\documentclass{rnaastex}
\usepackage{amsmath, amssymb, bm}

\begin{document}

\title{Ultra Diffuse Galaxies are a Subset of Cluster Dwarf Elliptical/Spheroidal Galaxies}

\correspondingauthor{Christopher J. Conselice}
\email{conselice@gmail.com}

\author{Christopher J. Conselice}
\affiliation{Centre for Astronomy and Particle Theory, University of Nottingham, Nottingham NG7 2RD, United Kingdom}

\keywords{galaxies: dwarf, galaxies: formation and evolution}


\vspace{-2cm}
\section*{Ultra Diffuse Galaxies are Previously Known} 

Since 2015 there has been a great deal of interest in a supposed new class of galaxy called Ultra Diffuse Galaxies (UDGs; van Dokkum et al. 2015).    These are large systems with sizes $> 1.5$ kpc and have surface brightness which are $\mu > 25$ mag arcsec$^{-2}$ (e.g., Koda et al. 2016).  Because of their low-surface brightness they are proposed to be `failed' Milky Way type galaxies given their similar size, but much lower stellar masses (e.g., van Dokkum et al. 2015).

As such, these systems are considered by some as a new type of galaxy, yet we show that they are a subset of a well-established population of low-surface brightness galaxies found mostly in dense areas of the universe - clusters of galaxies.  This population was discovered and reported in many previous papers (e.g., Sandage \& Binggeli 1984; Impey et al. 1988; Caldwell \& Bothun 1987) and in Conselice et al. (2003) where they were denoted as Low-Mass Cluster Galaxies (LMCGs).  The origin of LMCGs/UDGs may be different from dwarf ellipticals/dwarf spheroidals, but LMCGs/UDGs are similar in structure/morphological properties, and have a similar morphological and structural classifications.  There is also a continuum in properties from dwarf ellipticals/spheroidals to UDGs that suggest they are taken from the same population and differ only in size (e.g., Wittmann et al. 2017).

We show in this note that these galaxy `types' - the UDGs, LMCGs and cluster dwarf ellipticals, while perhaps not identical, have a large overlap and are essentially the same objects. To demonstrate this we show in Figure~1 the distribution of size and absolute magnitude for the LMCGs discovered in the Perseus cluster by Conselice et al. (2003), and those found by van Dokkum et al. (2015) in the Coma cluster.   As can be seen there is a large overlap in these two populations which differ only in name, although there are more larger-sized systems found in Coma.  This excess is due to the smaller area probed in the Perseus WIYN telescope survey than in the van Dokkum et al. (2015) Coma observations.   Furthermore, note that there is no correlation between the sizes of these systems and their magnitudes.  For most galaxies there is a scaling between these two, such that brighter systems are bigger (e.g., Norris et al. 2014).  This is a major clue to their formation method, which suggests an origin from the destruction of infalling galaxies, rather than failed construction.

It was also noted in the past that these systems must have a large dark matter content, given their large sizes and low surface brightness (Penny et al. 2009), to survive their host cluster tides.  This has later been confirmed with spectroscopy from e.g., van Dokkum et al. (2016).  However, the study by Penny et al. (2009) was the first to demonstrate, using reasonable arguments based on the strength of cluster tides, that these cluster LMCGs/dwarfs/UDGs must be dark matter dominated.

\begin{figure}[h]
\begin{center}
\includegraphics[scale=0.4,angle=0]{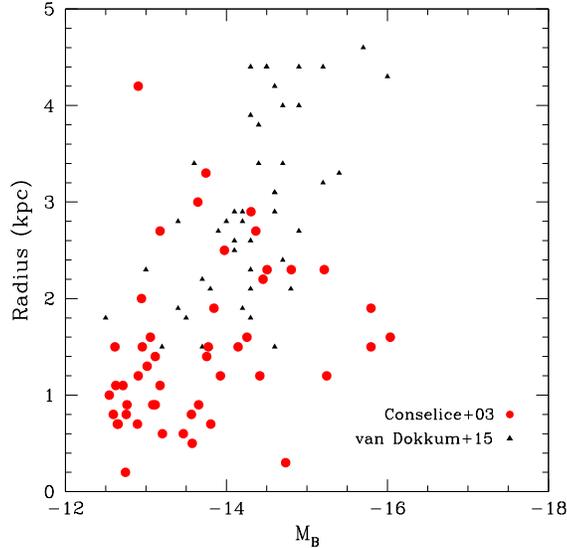}
\caption{Plot showing the Ultra Diffuse Galaxies (UDGs) from the Coma cluster
(black triangles) and the Low-Mass Cluster Dwarf Elliptial/Galaxies (red circle) from the similar rich Perseus cluster (Conselice et al. 2003), demonstrating the vast overlap between these two populations in different clusters.}
\end{center}
\end{figure}

\section*{Formation Mechanism}

It is thought because of their large sizes, but low surface brightness, that these systems somehow failed to form enough stars.  Multiple scenarios are proposed to account for this `failed' formation stage.    However, there are many reasons for believing that the bulk of the LMCG/UDG population was formed through cluster processes.  

Some indications for this are that these systems have a range of stellar population ages, consistent with being a mixture of very old and young (few Gyr) stellar populations with a mixture of metallicities, some being quite high (e.g., Penny \& Conselice 2008).  Also their distribution of radial velocities is non-Gaussian and has a much wider distribution than ellipticals which is similar to the infalling spirals (e.g., Conselice et al. 2001).  Furthermore, these systems are not often found outside of clusters and are most common in the densest areas of the universe (van der Burg et al. 2017).

The most likely method for forming these galaxies is through cluster processes such as `Galaxy Harassment' (e,.g., Moore et al. 1998), where through multiple high speed encounters an infalling galaxy is gradually removed of its mass, until it resembles a dwarf elliptical. As shown in detailed models of this process, an infalling spiral can lose a significant fraction of its mass and turn into what we would identify as a dwarf spheroidal/elliptical such as these UDGs (e.g., Mastropietro et al. 2005).

In conclusion, the Ultra-Diffuse galaxies are are previously discovered population that are a subset of the low-surface brightness dwarf galaxies which have been discovered and studied in clusters of galaxies for several decades.   As such, there is much to learn about these systems from the previous literature where they have been studied in depth for close to 20 years.  As described, they have a mixture of stellar populations, suggesting a diverse origin, which is further backed up by their kinematic distributions (e.g., Conselice et al. 2001; Penny \& Conselice 2008).  They also have a range of HI gas content, suggesting an origin from gas rich galaxies, having a pattern consistent with ram-pressure stripping (e.g., Conselice et al. 2003b; Buyle et al. 2005).

All of this evidence suggest that UDGs/LMCGs are a population in which at least a fraction arrived late into a cluster's environment.  It is much more likely that these systems formed from the destruction of infalling galaxies rather than the prevention of the formation of galaxies that failed to become more massive.  Future studies of UDGs should consider the above and their more general connection to previously studied populations.

\acknowledgments

I thank Sam Penny and Jay Gallagher for useful conversations and suggestions.





\end{document}